# A Flexible Infrastructure-Sharing 5G Network Architecture Based on Network Slicing and Roaming

João P. Ferreira [1], Vinicius C. Ferreira [2], Sérgio L. Nogueira [3], João M. Faria [3] and José A. Afonso [1,*]

[1] CMEMS-UMinho/LABBELS—Associate Laboratory, University of Minho, 4800-058 Guimarães, Portugal; pg47330@alunos.uminho.pt

[2] DTx – Digital Transformation CoLAB, University of Minho, 4800-058 Guimarães, Portugal; vinicius.ferreira@dtx-colab.pt

[3] dstelecom, S.A., 4700-727 Braga, Portugal; sergio.lopes@dstelecom.pt (S.L.N.); joaomanuel.faria@dstelecom.pt (J.M.F.)

* Correspondence: jose.afonso@dei.uminho.pt

**Abstract:** The sharing of mobile network infrastructure has become a key topic with the introduction of 5G due to the high costs of deploying such infrastructures, with neutral host models coupled with features such as network function virtualization (NFV) and network slicing emerging as viable solutions for the challenges in this area. With this in mind, this work presents the design, implementation, and test of a flexible infrastructure-sharing 5G network architecture capable of providing services to any type of client, whether an operator or not. The proposed architecture leverages 5G's network slicing for traffic isolation and compliance with the policies of different clients, with roaming employed for the authentication of users of operator clients. The proposed architecture was implemented and tested in a simulation environment using the UERANSIM and Open5GS open-source tools. Qualitative tests successfully validated the authentication and the traffic isolation features provided by the slices for the two types of clients. Results also demonstrate that the proposed architecture has a positive impact on the performance of the neutral host network infrastructure, achieving 61.8%-higher throughput and 96.8%-lower packet loss ratio (PLR) in a scenario sharing the infrastructure among four clients and eight users when compared to a single client with all the network resources.

## 1. Introduction

The introduction of 5G networks marks a significant revolution in the telecommunications world, providing multiple advantages over predecessor technologies such as reduced latency and higher data rates [1]. Designed with high reliability and accessibility in mind, these networks offer new functionalities to satisfy the exponential demand for services and device connectivity, embracing both personal communication and Internet of Things (IoT) application areas, including pervasive computing, healthcare, smart homes, smart cities, industrial automation, and vehicular networks [2].

One of the major drawbacks of 5G currently is the cost associated with deploying the new infrastructures required for the evolution and expansion of the mobile network. A potential solution to mitigate this issue is the implementation of neutral hosts as network service providers (NSPs) to enable infrastructure sharing among multiple tenants, the customer service providers (CSPs). The use of neutral hosts has been predominantly focused on rural areas, where deploying new networks might not be cost-effective for CSPs, as well as in public events and indoor spaces where there is limited room for each CSP to deploy individual infrastructures [3]. Another advantage of a 5G neutral host is the possibility of enabling non-mobile network operators (MNOs) to supply 5G services to their customers using the shared infrastructure.

From the CSPs' perspective, the reliability of these networks may not be optimal due to challenges in network management and commercial agreements, among other factors. The NSPs can use infrastructure virtualization to surpass those challenges, with techniques such as network slicing [4]. These advancements make neutral hosts more appealing and reliable; however, resource management and ensuring customer policy control are still challenges when using network slicing [5]. In this context, this paper proposes an infrastructure-sharing architecture that enables neutral hosts to provide their tenants a network slice with a set of 5G network services, being capable of offering CSPs the ability to provide a 5G service to their customers regardless of whether they are an MNO or not and ensure network services to provide control over their customers.

This paper's main contributions are as follows: the design of an infrastructure-sharing 5G network architecture that accommodates both types of tenants—MNOs and non-MNOs; the selection of a set of 5G network functions (NFs) to provide the tenants with all the necessary services, improving NSPs' scalability and resource management while ensuring CSPs' traffic isolation and policy control over their customers; the implementation of the proposed architecture in a simulation environment using open-source tools; and the execution of multiple tests to validate the proposed architecture.

The remainder of this article is organized as follows: Section 2 presents the background and related work. Section 3 describes the proposed solution and its implementation, whereas Section 4 presents the results and discussion. Finally, the conclusions and suggestions for future work are presented in Section 5.

## 2. Background and Related Work

This section presents a literature review of relevant background concepts that were used for the design of the proposed solution. Initially, an overview of the 5G network architecture is presented, followed by a description of various types of infrastructure sharing and neutral host models. Finally, an analysis of other works of significant interest to this paper is conducted.

*2.1. 5G Networks*

The 5G architecture is structured around four key components: the user equipment (UE), the radio access network (RAN), the transport network, and the core network (CN) [6]. The UE encompasses end-user devices, including mobile phones, IoT sensor devices, etc., whereas the transport network links the two critical modules at the heart of the 5G network: the CN and the RAN. The 5G CN, or 5GC, is the central part of the 5G mobile network architecture, responsible for managing mobile voice, data, and Internet services.

The Third-Generation Partnership Project (3GPP) has delineated the 5GC architecture as a service-based architecture (SBA) composed of several modular entities known as network functions (NFs) [5]. The NFs communicate with each other using standardized interfaces, which allows for interoperability and flexibility in network deployment [4]. The 5GC NFs are exclusively software-based and are designed for cloud deployment, obviating the need for dedicated hardware for each function and enhancing deployment agility and flexibility.

The main NFs as described by 3GPP are [5] as follows: access and mobility management function (AMF), responsible for registration, authentication, and tracking of the UE; session management function (SMF), which establishes and manages sessions for user data, supports customized mobility management schemes together with AMF, coordinates data routing and forwarding with the UPF, and is responsible for policy enforcement and QoS; and user plane function (UPF), which connects the mobile infrastructure and the data network (DN), being responsible for packet routing and forwarding, packet inspection, and QoS handling for the user plane. Other relevant functions that are used in this work are as follows: network repository function (NRF), which supports NF services management including registration, deregistration, authorization, and discovery; policy control function (PCF), which supports unified policy framework to govern network behavior and provides policy rules to control plane functions; and security edge protection proxy (SEPP), which supports message filtering and policing between service consumers on public land mobile network (PLMN) control plane interfaces.

The RAN is responsible for managing the radio spectrum and ensuring the quality of service (QoS) for the end-user through implementing the 5G base stations, known as gNodeB (gNB), which provide wireless connectivity to the end-user devices. The gNB tasks relevant to this work are the selection of an AMF at UE attachment when no routing to an AMF can be determined from the information provided by the UE, the routing of user plane data towards the UPF(s), the routing of control plane information towards the AMF, and the support of network slicing (NS) [5].

*2.2. Network Slicing*

The concept of network slicing involves the creation of configurable and independent virtual networks within the same physical network infrastructure. Each network slice can be tailored with dedicated resources, meeting the specific needs of the service it aims to provide. Virtualization plays a pivotal role in enabling NS, allowing for the dynamic and efficient allocation of physical infrastructure resources and ensuring individualized traffic isolation within each network slice [7]. In this sense, within the realm of 5G, NS entails the amalgamation of necessary physical resources and virtual network functions (VNFs) to deliver a distinct service, isolated from others while sharing the same physical infrastructures. A network slice encompasses both control and user plane network functions within the CN, as well as the RAN, and is defined within the domain of a PLMN. Network slice identification is achieved through single-network slice selection assistance information (S-NSSAI), comprising a slice/service type (SST) and a slice differentiator (SD) [8].

*2.3. 5G Roaming*

Roaming refers to agreements between operators that allow for the expansion of a home PLMN (HPLMN) service coverage into areas where it is not present. This enables subscribers to use the services of another operator, whether in another country (international roaming) or within their own country (national roaming). Two types of roaming architecture supported in 5G are identified [9]:

- Local breakout (LBO): In this model, the user's authentication and authorization on the visited network are performed by the home network, while the traffic is directly routed through the visited PLMN (VPLMN).
- Home routed (HR): In this scenario, the data traffic from the VPLMN is routed to the data network (DN) via the HPLMN.

For communication between the HPLMN and the VPLMN, the Third-Generation Partnership Project (3GPP) defined a new network function: the security edge protection proxy (SEPP), which acts as a proxy between PLMNs to provide security and privacy to the involved entities. All control plane messages pass through the SEPP of both the home and visited networks, ensuring the protection of messages before forwarding them to network functions or the service communication proxy (SCP).

The SCP was introduced by the 3GPP in the 5G system to enable indirect communication between NFs. In addition to performing load balancing, the SCP also handles overload. The SEPP can forward messages directly to the NFs or opt to use the SCP. When used, the SCP supports inter-PLMN routing, providing the necessary means for forwarding messages relevant to the SEPP [9].

*2.4. Neutral Hosts*

Neutral host models enable network sharing among multiple tenants under the management of MNOs or third-party business organizations, ensuring connectivity and coverage for devices across public and/or private mobile network operators.

In the context of 5G, neutral host networks provide a wide array of applications and benefits. Many use-cases for these networks are concentrated in rural areas, university campuses, public events, and indoor spaces. In rural areas, the cost of implementing individual networks for each MNO is not justifiable due to low population density, making neutral host networks a suitable solution to minimize costs for all parties involved. In indoor areas, university campuses, and public events, the reduction in the space needed for infrastructure deployment, cost, and energy consumption makes infrastructure sharing an ideal solution [7].

This approach also enables MNOs to expand their mobile network coverage, addressing the increasing difficulty in acquiring sites for antennae and base stations, facilitating a swifter evolution of 5G, and reducing the cost and development time of technologies to meet the functionalities and demands of 5G [3,7].

However, neutral host models also present disadvantages and pose challenges in their development. One major obstacle is the loss of end-to-end network visibility by operators, leading to reliance on third-party companies for network maintenance. This can result in diminished control over service quality and customer experience. As a result, service-level agreements (SLAs) would become more demanding and complex [7,10]. Additionally, the lack of dedicated resources and software is a challenge as networks are not planned and designed with neutral host models in mind. Security and implementation are also concerns.

Figure 1 illustrates different models for sharing 5G network infrastructures, categorizing them into two groups: passive sharing and active sharing, where A and B represent two different MNOs [11]. Passive sharing involves the sharing of non-electronic infrastructure components, such as sites and towers, power supplies, and the physical elements of transport in backhaul, e.g., optical fibers. Site sharing is a common and straightforward practice that allows operators to maintain their competitive strategies. It is operationally simple since the equipment is independent for each operator, yet it can still provide long-term cost reductions. Active sharing refers to the sharing of electronic components of the network, including elements of the RAN, such as base stations, antennae, controllers, etc. It can also encompass spectrum sharing and even sharing the CN, including servers and network functionalities.

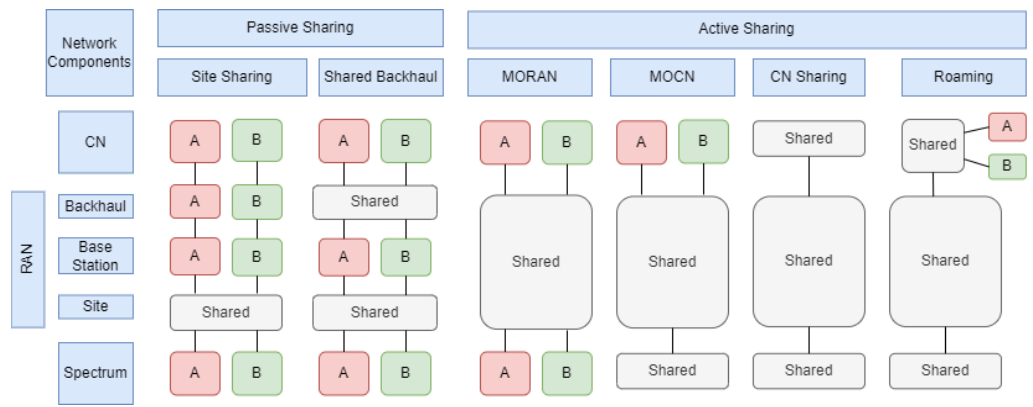

**Figure 1.** Classification of infrastructure sharing models (adapted from [11]).

Regarding RAN sharing, there are two main models: the multiple-operator core network (MOCN) and the multiple-operator radio access network (MORAN). The MOCN involves sharing the RAN and spectrum by multiple operators, whereas in the MORAN, there is no spectrum sharing, so operators must use their dedicated spectrum range. CN sharing is much less common due to the complexity of operation and maintaining strategy differentiation among operators, making it a less attractive model despite being more economical than RAN sharing.

Roaming is also a form of infrastructure sharing. In the case of a neutral host, when an operator lacks coverage in an area, it can use another operator's coverage through a roaming agreement [9,11].

*2.5. Related Work*

Network slicing is a technique that may be used for different purposes, including a plethora of 5G scenarios at different network domains (RAN, CN, transport, general) and even with non-terrestrial networks [8,12]. The use of slicing in the RAN domain poses resource management challenges, especially in spectrum time and spatial domain, with multi-tiered and AI-supported approaches being promising solutions [6]. Nevertheless, these solutions are out of the scope of this paper, which focuses on neutral host or multi-tenant terrestrial mobile network architecture proposals and does not take into account the spectrum sharing problem.

Kassis et al. [13] proposed a RAN-sharing model based on a flexible network slicing mechanism that leverages the efficiency of MOCN along with the spectrum isolation inherent in MORAN through a dynamic transition between these two scenarios. The implementation was based on open-source 4G software from OpenAirInterface (OAI) and used an SDN approach to manage the separation of control and user planes in the MAC layer, using the FlexRAN RAN controller provided by the Mosaic5G project, which facilitates the real-time creation and configuration of network slices. The authors also proposed an algorithm for dynamic radio resource allocation for each slice. In their tests, they compared their approach to a static RAN sharing scenario and concluded that their architecture allows for efficient radio resource management, considering factors like traffic demand, SLAs, and required isolation. Although the authors present an innovative proposal for RAN sharing among multiple operators, some potential challenges are the implementation complexity, the need for testing in diverse scenarios, and the technically complex management requirements.

In [14], the author proposed a neutral host architecture for indoor spaces where the neutral host has complete control over the network infrastructures and elements, and the interaction with service providers (SPs) is established through roaming agreements. Using the Citizen Broadband Radio Service (CBRS) spectrum, in the 3.5 GHz band, which is available for shared commercial use in the United States, the architecture allows a single small cell to serve multiple operators. To maintain the visibility and trust of the SPs in the neutral host network, trust relationships between entities were established, along with providing detailed information about network performance quality through key performance indicators (KPIs). This approach is similar to the one used in this paper, as both employ roaming to authenticate users. However, the author does not consider the isolation and separation of the traffic among SPs nor, consequently, ways to ensure and maintain the desired QoS as agreed in the contracts with the SPs.

Giannoulakis et al. [15] presented the architecture of the 5G SESAME project of the 5G public–private partnership (5G-PPP), which is composed of two main components—the cloud-enabled small cell (CESC) and the CESC manager (CESCM)—and leverages SDN and NFV technologies to enable small cells to support multiple operators through the request of network slices from the infrastructure. The SESAME architecture envisages the separation of physical and virtual functions based on the MOCN model.

The 5G-City project [10], which is also a 5G-PPP initiative, proposes an architecture split vertically across three layers: service/application layer; orchestration and control layer; and infrastructure layer. The proposed architecture was deployed in three different cities, Barcelona, Bristol, and Lucca, and the feasibility of deploying a multi-tenant virtualized infrastructure was demonstrated.

Another 5G-PPP initiative, the 5G-ESSENCE project [16], presents an interesting approach to edge network deployment and infrastructure sharing. It uses a two-tier system, with edge data centers for low-latency services and small cell virtualization and a central data center for heavy processing to support public safety applications, offering priority to first responders and emergencies. Each supported public safety service was associated with a different network slice, and the project was validated using an LTE infrastructure.

This paper's proposal aims at solving the 5G infrastructure sharing problem by using roaming and network slicing. Our neutral host architecture embraces two different types of tenants: an MNO type, which receives only the necessary NFs to authenticate and control its customer QoS; and a non-MNO type, which receives all NFs to provide a 5G service to its users. This approach keeps the visibility and trust of the SPs, isolates and separates traffic among SPs, and reduces the complexity and management by providing only the necessary NFs to each SP according to its needs and agreements.

## 3. Proposed Architecture

The development of a 5G neutral host architecture plays a pivotal role in the evolution of a 5G network into a service-oriented format. This section presents a detailed description of the proposed architecture and its implementation using open-source software, outlining the selected software and the decision-making process and highlighting the choices and limitations involved.

### 3.1. Conceptual Architecture

The proposed neutral host architecture aims to promote efficient sharing of 5G network resources and infrastructures among different clients. For this purpose, it provides service without the need for third-party network elements in the connection and routing of traffic while also allowing for any type of CSP, whether MNOs or not.

To distinguish between different types of stakeholders throughout the paper, CSPs who are MNOs have been termed "operator clients", while those who are not MNOs are referred to as "non-operator clients". The NSP, which provides the shared infrastructure, is named "neutral host". Lastly, these clients may support various UEs corresponding to their service subscribers, who are referred to as "users".

The proposed architecture, exemplified in Figure 2, leverages network slicing and RAN sharing with the LBO roaming architecture described in Section 2.3. This figure illustrates three distinct network slices, each represented by a different color and dedicated to a specific client. As represented, while all UEs access the network through the shared RAN and the access and mobility management function (AMF), indicated by solid lines, each client has its dedicated slice, responsible for managing the session, along with the user plane function (UPF), session management function (SMF), and policy control function (PCF). In this example, two operator clients (Slice 2 and Slice 3), with their respective PLMN IDs 2 and 3, are requesting service via roaming agreements, while the non-operator client (Slice 1), with PLMN ID 1, does not require roaming agreements and is authenticated directly in the neutral host's core.

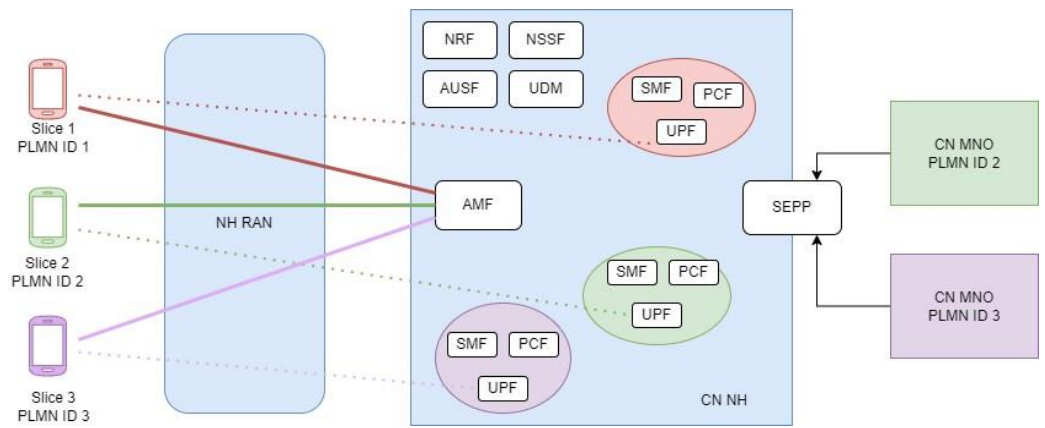

**Figure 2.** Conceptual neutral host architecture of the proposed solution.

Figure 3 represents the procedures undertaken to enable a new client on the neutral host network. The steps are slightly different depending on whether the client is an operator or not. For an operator client, we use roaming agreements to provide user authentication and set up the neutral host SEPP; whereas for a non-operator client, it is necessary to obtain a user database and set UDM and AUSF to perform the user's authentication. Given those different users and authentication methods, the user's initial attach registration procedure also differs, i.e., from a roaming authentication to a direct NH authentication, if it is from an operator or non-operator client, respectively, as illustrated in Figure 4.

Typically, the predominant models proposed for neutral hosts focus either on passive sharing or RAN sharing with the implementation of MORAN or MOCN functionalities. These approaches have been the most common among MNOs for cost savings to date. However, the advent of 5G and emerging technologies, such as NFV, SDN, and RAN centralization, broaden the operators' perspectives with respect to new types of infrastructure sharing, enabling MNOs to maintain network control even when operating on third-party infrastructures. Therefore, RAN sharing is a component of the proposed architecture, as it is necessary to share the same RAN among different clients and, consequently, different PLMN IDs. However, unlike the MORAN and MOCN models, this architecture does not route traffic to multiple cores, containing only the core of the neutral host and catering not just to operators but also to non-operator clients.

The proposed architecture aims for core sharing and leverages 5G-associated network slicing to provide an isolated and secure service for each client. This proposal involves implementing a core with all NFs but granting the client a network slice composed of three NFs, namely, SMF, PCF, and UPF, which directly impact the user experience and are crucial in controlling QoS and configuring user sessions. Thus, this architecture offers flexibility and dynamism to the network, allowing clients to configure QoS policies and maintain visibility over their network slice. It also ensures isolation and traffic control for each client, ensuring that the traffic and policies of one slice do not affect others. To meet these requirements, it is essential to have agreements between the clients and the neutral host. Implementing techniques to share network performance indicators, monitoring tools, and application programming interfaces (APIs) is crucial to ensure network visibility for clients.

In this architecture, roaming plays a crucial role in the authentication and verification of users belonging to operator clients. The LBO model, chosen in our approach, allows for user traffic to be locally routed to the Internet, unlike the HR model, where traffic is directed to the HPLMN. Despite the latter being seen as maintaining more visibility over the traffic for the MNOs, by providing a specific network slice for each client, together with an effective approach in terms of APIs and performance indicators availability, the proposed architecture allows them to maintain visibility and control over their users, QoS policies, and services. Roaming is fundamentally involved only in terms of authentication and authorization. To maintain control over their users, the operator clients themselves ensure the authentication within the neutral host's network, thus leveraging the standardized security and reliability for roaming.

For non-operator clients' users, authentication is performed on the neutral host's network as is typical in traditional mobile networks, given that they do not require roaming agreements. For operator clients in roaming service, the rationale is to allow the neutral host to use the same PLMN ID used by the operator client's mobile network to provide connectivity to its users. This approach simplifies the switching process between the VPLMN (neutral host network) and HPLMN (home network) for the user's device, as the devices maintain a constant connection to the same PLMN ID. This ensures faster and more transparent network switching for the user, unlike what typically occurs in international roaming, where prolonged service losses and roaming notifications are common, consequently improving the user experience when using the neutral host network.

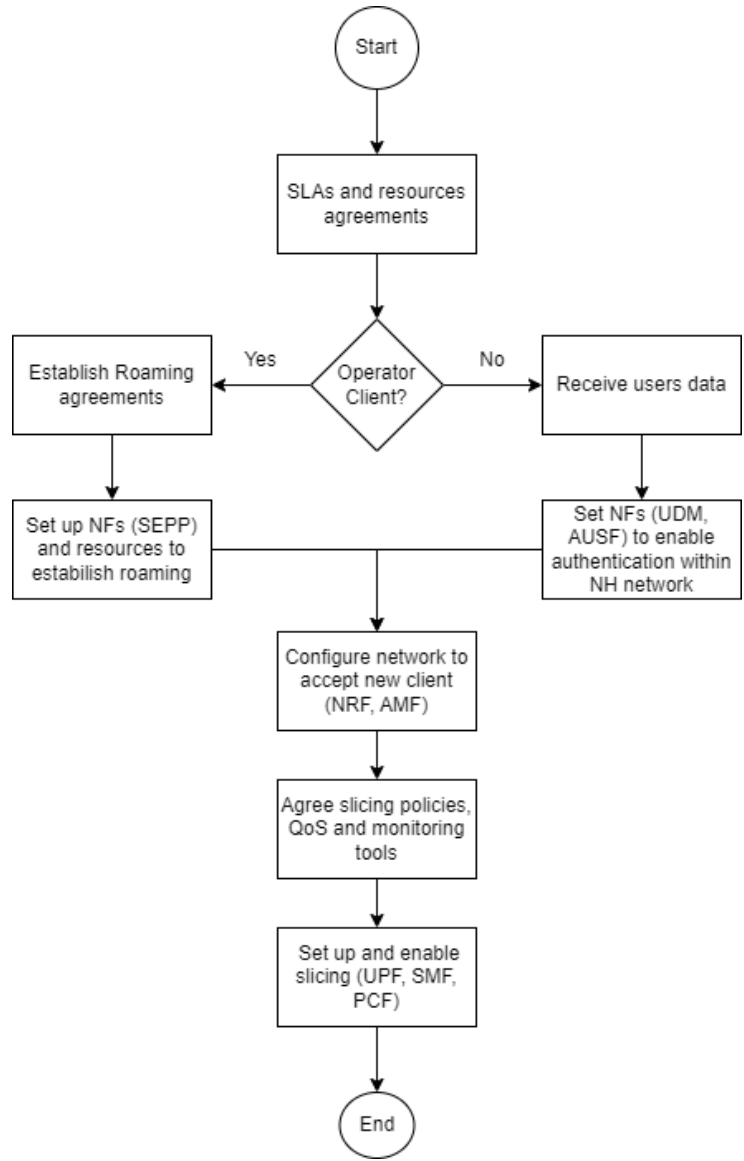

**Figure 3.** Flowchart of a new client configuration in the neutral host.

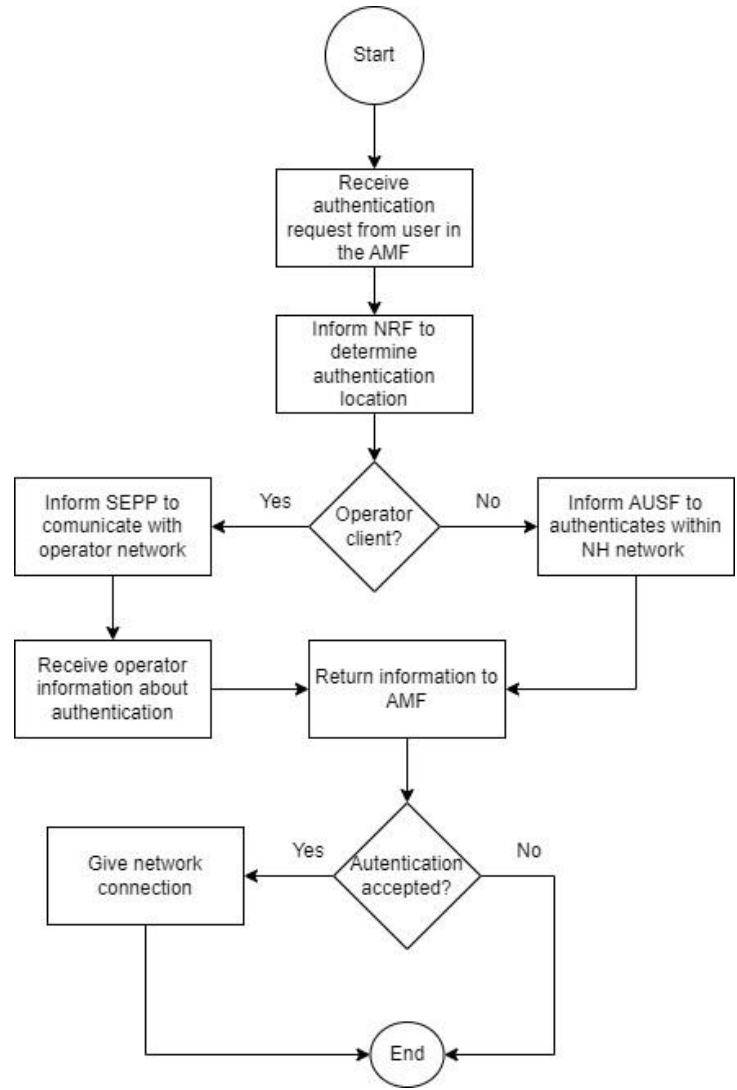

**Figure 4.** Flowchart of the authentication procedure.

*3.2. Architecture Implementation*

3.2.1. Software Tools

In this work, several open-source software tools were considered for implementation and testing of the proposed architecture. For the RAN simulation, notable options include UERANSIM [17], OpenAirInterface [18], and my5G-RANTester [19], while for the CN, Open5GS [20], free5GC [21], and Openair-cn [22] are prominent.

In the context of RAN simulation, UERANSIM was the chosen option. Although OpenAirInterface offers more features, its usage demands significantly higher computational capacity and is also more complex to configure and use. Moreover, OpenAirInterface was unable to connect more than one UE to a gNB on the same virtual machine (VM), which brought the necessity to use more VMs and, consequently, higher computer resource demands. On the other hand, both UERANSIM and my5G-RANTester were found to be simpler to use and configure and are capable of running on VMs with lower computational capacity. The choice of the former was based on the considerably higher availability of information on the Internet about it, its use in various projects, and the clear separation between the UE and the gNB, unlike my5G-RANTester, which integrates both in the same terminal. Table 1 illustrates the most significant features taken into account in the choice of RAN simulator software for use in this work.

Table 1. Characteristics of the RAN simulator alternatives.

| RAN Simulator | Multiple PLMNs | Computational Resources | Multiple UEs in the Same VM | Minimal CPU Core Capacity Per VM |
|---|---|---|---|---|
| **OpenAirInterface** | Yes | High | No | 4 |
| **UERANSIM** | No | Low | Yes | 1 |
| **My-5GRANTester** | No | Low | Yes | 1 |

In the context of the CN simulation, Open5GS was selected. Reddy et al. [23] conducted a study comparing the features of open-source software for simulating the 5G CN. Regarding openair-cn, they concluded that it requires greater CPU capacity than the other options. Concerning free5GC and Open5GS, the authors concluded that both are easy to configure, use, and integrate with UERANSIM. However, the selection of Open5GS was based on several key points. After testing both simulators, it was found that free5GC had limitations in handling different PLMN IDs. Although it correctly accepted and interpreted multiple PLMN IDs in the configuration, it failed to authenticate different UEs with different PLMN IDs. Furthermore, Open5GS has recently included support for roaming, which is crucial to this work and not available in the other two options.

3.2.2. Implemented System

Taking into account the chosen software tools and computational resource limitations, a basic scenario was developed to implement and validate the proposed solution. According to the proposed conceptual architecture (Figure 2), the RAN should support multiple PLMNs to simulate their sharing. However, since the UERANSIM RAN simulator does not have this capability, we virtualized an instance of UERANSIM for each client in a different VM.

Regarding the CNs, compared to the conceptual architecture proposed, the interfaces implemented by Open5GS are limited to the authentication of the UE. This means that only the AMF interacts with the operator's CN for UE authentication and that it does not include the necessary connections between the SMF and this CN for updating and verifying authentications. Regarding the slicing envisaged in the proposed conceptual architecture, there were also some changes in the implemented architecture due to some limitations of Open5GS. Therefore, only the NFs SMF and UPF are part of each client's slice. Finally, issues of billing policies, resource distribution, and QoS management are not addressed in this paper. In the current architecture implementation, each slicing is manually configured using the functionalities provided by Open5GS. This tool supports slicing and NFs creation by changing the configuration files of Open5GS, and a service restart is required to apply the new settings and NFs.

Figure 5 illustrates the simplified implemented architecture, which consists of different instances of VMs wherein each client has a dedicated UERANSIM instance within a dedicated VM, and their number is contingent on the number of clients. The core networks of the operators were implemented in the same VM as the neutral host core, as the Open5GS does not require their separation.

To demonstrate the development of the implemented architecture, highlighting the configuration of virtual machines and the software used in interactions between different parts of the system, a basic scenario was designed. This scenario considers both types of clients: operator and non-operator. Figure 5 presents the implemented system, as well as the VMs used and the interactions between them.

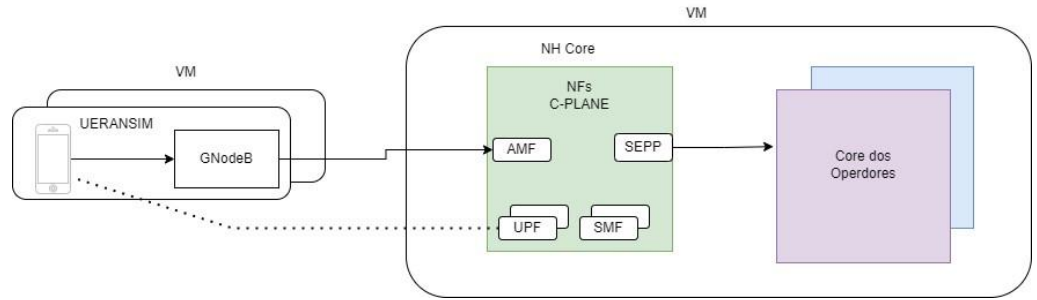

**Figure 5.** Architecture of the implemented system.

Figure 6 presents a more detailed illustration of the implemented architecture, emphasizing the used VMs and their interactions. This scenario, which was used on the tests presented in Section 4, includes up to four clients: one operator client and up to three non-operator clients, as well as four SMFs and UPFs. This figure represents the most comprehensive architecture employed in the tests, the places where the UERANSIM VMs were used, and the SMF-UPF pairs for each slice, which are proportional to the number of clients.

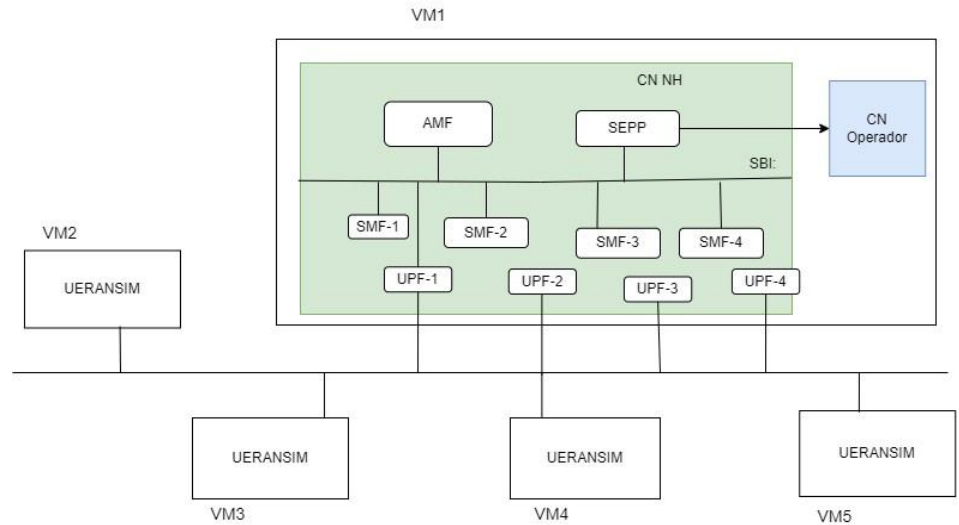

**Figure 6.** Architecture of the evaluation scenario used for the tests.

All the VMs used run the Ubuntu 20.04 operating system. The VM with the neutral host core has two CPU cores, 2 GB of RAM, and 20 GB of storage, while the VMs with UERANSIM are equipped with one CPU core, 1 GB of RAM, and 15 GB of storage.

## 4. Results and Discussion

This section describes the tests conducted to validate the architecture proposed in this paper, the results obtained, and the corresponding discussion. The results are split into three sections. The first one involves functional testing to check the MNO and non-MNO authentication methods and the traffic isolation of a network slice. Afterward, we analyze our proposal's efficiency to share and manage the NSP resources by measuring the impact on CSPs and their users. These performance evaluation results are split into two sections, one using the throughput and the other using the packet loss as metrics. In those tests, we use the maximum number of slices given our hardware limitations, which are four slices, and use as the baseline an NSP with a single client, which corresponds to having a single network slice with all the available resources. The performance evaluation results include the confidence intervals for a confidence level of 95%.

*4.1. Authentication and Traffic Isolation*

In order to validate the implemented architecture, the user authentication procedure and the network slice traffic isolation were tested for both types of clients: operator and non-operator. A simple test scenario with the infrastructure shared by one operator client and one non-operator client was designed, and we measured the user authentication time and verified the traffic isolation among clients.

Regarding the authentication time for the two types of clients, 10 measurements were taken from a user of each client, the averages were calculated, and the difference was analyzed. The average value obtained for the authentication of the operator-client user was 262.4 ms, and for the non-operator-client user, it was 239.9 ms. Thus, a slight difference in the authentication time of 22.5 ms was observed. This difference is a consequence of the extra network hops in the authentication of the operator-client user as it requires the core of its operator to authenticate it via roaming, while the non-operator-client user authenticates directly on the neutral host CN. This time difference can become more significant according to the communication delay between the operator CN and the neutral host CN.

Open5GS creates a tunnel interface for each slice, so to verify traffic isolation in each slice, traffic was generated using iperf from two different users of different clients. On the core network side, tcpdump was used on each interface to monitor the traffic of each user. Through filters, we checked if the traffic passing through one user's interface also passed through the other user's interface. If it happened, it would indicate interference between the interfaces of each user. However, we observed that this was not the case, leading us to conclude that each interface is specifically dedicated to each client, thus ensuring traffic isolation between the slices.

*4.2. Throughput*

This test, which used the evaluation scenario presented in Figure 6, measured the maximum throughput obtained in the utilization of the infrastructure and compared using our neutral host architecture, sharing resources with a network slice per different clients, and using the neutral host infrastructure as a traditional network operator, i.e., one client/one slice serving its users. Given the resource limitations of our testing hardware, we limited the maximum number of clients to test in our proposed architecture to four, resulting in four network slices. To overcome the limitations imposed by the VMs with UERANSIM, and to make it a fair comparison between the two configurations, the users were distributed across the same number of different UERANSIM VMs both in the single client configuration and in the configuration with four clients.

The user traffic was generated using the iperf tool [24] over TCP (transmission control protocol). The maximum throughput per user achieved with each configuration was measured in periods of 60 s and with the parameter value of window size set to 416 kB. The values shown in the tables correspond to an average of 10 recorded values. In the tables, the different IP (Internet protocol) addresses identify the different clients present in the architecture.

Table 2 represents the maximum throughput per user and total throughput with eight users, illustrating the results obtained with the two different configurations described before. In the first row, the eight users belong to the same client and are evenly distributed across four VMs, whereas in the second row, the eight users are evenly distributed across four different clients/slices and their respective VMs. Table 3 presents results from similar tests, but comprising only four users, one per each UERANSIM VM.

Table 2. Maximum throughput with eight users distributed by four VMs for both configurations.

| | Throughput (Mbps) | | | | | | | | TOTAL |
|---|---|---|---|---|---|---|---|---|---|
| | VM2 | | VM3 | | VM4 | | VM5 | | |
| One client/slice | 22.0 ± 2.7 | 23.6 ± 2.9 | 21.0 ± 2.9 | 22.1 ± 2.9 | 22.4 ± 2.3 | 22.1 ± 2.5 | 21.7 ± 3.3 | 22.1 ± 2.3 | 177.2 |
| Four clients/slices | 31.5 ± 3.5 | 35.9 ± 5.1 | 36.4 ± 5.3 | 33.2 ± 4.4 | 38.3 ± 5.3 | 37.5 ± 3.2 | 38.1 ± 3.2 | 35.7 ± 3.6 | 286.7 |

Table 3. Maximum throughput with four users, one per VM, for both configurations.

| | Throughput (Mbps) | | | | TOTAL |
|---|---|---|---|---|---|
| | VM2 | VM3 | VM4 | VM5 | |
| One client/slice | 49.8 ± 3.2 | 50.7 ± 4.1 | 51.1 ± 3.3 | 49.8 ± 3.2 | 201.4 |
| Four clients/slices | 89.0 ± 13.0 | 66.6 ± 11.6 | 89.7 ± 11.9 | 80.3 ± 12.8 | 325.7 |

The results in Table 2 show a clear increase in the throughput of the eight users—and, consequently, in the total throughput—for the configuration with four clients/slices compared to the one with only one client/slice, both cases with two users per VM. The same pattern can be observed in Table 3, with one user per VM. The increase in the total throughput was similar (61.8% in the first table and 61.7% in the second one). These results show that the proposed neutral host architecture has a positive impact on the user throughput.

Comparing the results for the same configuration from both tables, we can see that for both configurations, the total throughput is higher when there are only four users in the neutral host network instead of eight. This is expected since in the first case, two users are competing for the processing resources of each UERANSIM VM compared with one per VM in the second case.

To better understand these results, the CPU usage was monitored during these experiments in the VM1, the neutral host CN. It was observed that in all cases, an individual process was formed for each slice. In the configurations with only one slice, the CPU usage increased as the number of users increased but remained limited to a single process, resulting in a maximum CPU usage that did not exceed 99%. In contrast, in the configurations with four slices, four distinct processes were generated, one for each slice. This approach allowed for a more efficient distribution of traffic among the different slices, resulting in individualized CPU usage. As a result, the CPU usage was not restricted to a single core but was distributed among the available cores in VM1. This led to an increase in maximum throughput per user.

*4.3. Packet Loss Ratio*

This section demonstrates the behavior of the neutral host network in terms of the packet loss of different clients. To measure the packet loss ratio, the user traffic generated by the iperf tool, in this case, was over UDP (user datagram protocol) since it does not recover from lost packets as TCP does. The throughput per user was selected based on the maximum feasible value that allowed us to keep all eight users connected. The loss of connection for high loads that was observed while conducting these tests (and even failure to connect in some cases) can be justified by limitations of the simulator and the available computational resources.

Table 4 presents the packet loss ratio (PLR) experienced by the user connections with the maximum feasible throughput per user. Except for the fixed throughput per user and the use of UDP traffic, the configurations are similar to those used in Table 2.

Table 4. PLR and maximum feasible throughput per user with eight users distributed by four VMs for both configurations.

|  | PLR (%) | | | | | | | |
| --- | --- | --- | --- | --- | --- | --- | --- | --- |
|  | VM2 | | VM3 | | VM4 | | VM5 | |
| **One client/slice** | 1.11 ± 0.37 | 1.12 ± 0.38 | 1.10 ± 0.39 | 1.13 ± 0.41 | 1.06 ± 0.45 | 1.05 ± 0.44 | 1.19 ± 0.41 | 1.15 ± 0.39 |
| **Four clients/slices** | 0.05 ± 0.05 | 0.04 ± 0.03 | 0.03 ± 0.03 | 0.03 ± 0.02 | 0.04 ± 0.04 | 0.04 ± 0.04 | 0.03 ± 0.02 | 0.03 ± 0.02 |

According to these results, the proposed neutral host architecture experienced a mean PLR 96.8% lower than the configuration with just one client (0.036%, on average, compared to 1.11%). To ensure a fair comparison, both configurations were tested with a throughput of 23.1 Mbps.

Thus, similarly to what was observed with the throughput, the differentiation of processes for each slice leads to better CPU distribution in the neutral host CN VM, resulting in a positive impact on network performance in terms of packet loss per user compared to using only one slice.

## 5. Conclusions

This paper explored the sharing of 5G infrastructure through the development of a flexible and dynamic architecture that allows for the efficient provision of mobile network services to various types of clients, regardless of whether they are operators or not, while ensuring an efficient provision of the services and the isolation of each client's traffic.

The complexity of the proposed architecture, coupled with limitations of the available open-source software tools and computational resources, led to some restrictions in the architecture implemented for validation of the proposed solution.

Nevertheless, the tests performed in this work using the implemented neutral host architecture showed promising results, with successful validation of the different types of authentication and traffic isolation requirements, and a significantly better performance (higher throughput and lower PLR per user) of the proposed architecture based on network slicing and roaming when compared to the baseline test, i.e., a single client using all the available network resources.

**References**

1. Jabagi, N.; Park, A.; Kietzmann, J. The 5G Revolution: Expectations Versus Reality. *IT Prof.* **2020**, *22*, 8–15. https://doi.org/10.1109/MITP.2020.2972139.
2. 5G PPP Architecture Working Group. *View on 5G Architecture*; Version 4.0; 5G PPP Architecture Working Group: Pisa, Italy, 2021.
3. Kumar, S.K.A.; Crawford, D.; Stewart, R. Pricing Models for 5G Multi-Tenancy Using Game Theory Framework. *IEEE Commun. Mag.* **2023**, 1–7. https://doi.org/10.1109/MCOM.001.2200742.
4. Baldoni, G.; Cruschelli, P.; Paolino, M.; Meixner, C.C.; Albanese, A.; Papageorgiou, A.; Khalili, H.; Siddiqui, S.; Simeonidou, D. Edge Computing Enhancements in an NFV-Based Ecosystem for 5G Neutral Hosts. In Proceedings of the 2018 IEEE Conference on Network Function Virtualization and Software Defined Networks (NFV-SDN), Verona, Italy, 27–29 November 2018; pp. 1–5.
5. 3rd Generation Partnership Project. *3GPP Technical Report 21.915*; Version 15.0.0 Release 15; ETSI: Sophia Antipolis, France 2019.
6. Zhou, C.; Gao, J.; Li, M.; Shen, X.; Zhuang, W.; Li, X.; Shi, W. AI-Assisted Slicing-Based Resource Management for Two-Tier Radio Access Networks. *IEEE Trans. Cogn. Commun. Netw.* **2023**, *9*, 1691–1706. https://doi.org/10.1109/TCCN.2023.3307929.
7. Bajracharya, R.; Shrestha, R.; Jung, H.; Shin, H. Neutral Host Technology: The Future of Mobile Network Operators. *IEEE Access* **2022**, *10*, 99221–99234. https://doi.org/10.1109/ACCESS.2022.3207823.
8. Chahbar, M.; Diaz, G.; Dandoush, A.; Cérin, C.; Ghoumid, K. A Comprehensive Survey on the E2E 5G Network Slicing Model. *IEEE Trans. Netw. Serv. Manag.* **2021**, *18*, 49–62. https://doi.org/10.1109/TNSM.2020.3044626.
9. Keller, R.; Castellanos, D.; Sander, A.; Robison, A.; Abtin, A. Roaming in the 5G system: The 5GS roaming architecture. *Ericsson Technol. Rev.* **2021**, *2021*, 2–11.
10. Fernández-Fernández, A.; Colman-Meixner, C.; Ochoa-Aday, L.; Betzler, A.; Khalili, H.; Siddiqui, M.S.; Carrozzo, G.; Figuerola, S.; Nejabati, R.; Simeonidou, D. Validating a 5G-Enabled Neutral Host Framework in City-Wide Deployments. *Sensors* **2021**, *21*, 8103. https://doi.org/10.3390/s21238103.
11. GSMA. Infrastructure Sharing. Available online: https://www.gsma.com/futurenetworks/wiki/infrastructure-sharing-an-overview/ (accessed on 23 February 2024).

12. Esmat, H.H.; Lorenzo, B.; Shi, W. Toward Resilient Network Slicing for Satellite–Terrestrial Edge Computing IoT. *IEEE Internet Things J.* **2023**, *10*, 14621–14645. https://doi.org/10.1109/JIOT.2023.3277466.
13. Kassis, M.; Costanzo, S.; Yassin, M. Flexible Multi-Operator RAN Sharing: Experimentation and Validation Using Open Source 4G/5G Prototype. In Proceedings of the 2021 Joint European Conference on Networks and Communications & 6G Summit (EuCNC/6G Summit), Porto, Portugal, 8–11 June 2021; pp. 205–210.
14. Mun, K. *Making Neutral Host a Reality with OnGo*; Mobile Experts White Paper; Mobile Experts, Inc.: Campbell, CA, USA, 2018.
15. Giannoulakis, I.; Xylouris, G.; Kafetzakis, E.; Kourtis, A.; Fajardo, J.O.; Khodashenas, P.S.; Albanese, A.; Mouratidis, H.; Vassilakis, V. System Architecture and Deployment Scenarios for SESAME: Small cEllS coodinAtion for Multi-Tenancy and Edge Services. In Proceedings of the 2016 IEEE NetSoft Conference and Workshops (NetSoft), Seoul, Republic of Korea, 6–10 June 2016; pp. 447–452.
16. Spada, M.R.; Pérez-Romero, J.; Sanchoyerto, A.; Solozabal, R.; Kourtis, M.A.; Riccobene, V. Management of Mission Critical Public Safety Applications: The 5G ESSENCE Project. In Proceedings of the 2019 European Conference on Networks and Communications (EuCNC), Valencia, Spain, 8–21 June 2019; pp. 155–160.
17. Güngör, A. UERANSIM. Available online: https://github.com/aligungr/UERANSIM (accessed on 16 October 2023).
18. OpenAirInterface. OAI 5G RAN Project Group. Available online: https://openairinterface.org/oai-5g-ran-project/ (accessed on 16 October 2023).
19. my5G. my5G-RANTester. Available online: https://github.com/my5G/my5G-RANTester (accessed on 16 October 2023).
20. Open5GS. Available online: https://open5gs.org/ (accessed on 16 October 2023).
21. free5GC. Available online: https://free5gc.org/ (accessed on 16 October 2023).
22. OpenAirInterface. 5G Core Network. Available online: https://openairinterface.org/oai-5g-core-network-project/ (accessed on 16 October 2023).
23. Reddy, R.; Lipps, C.; Schotten, H.D.; Gundall, M.; Schotten, H.D. Open Source 5G Core Network Implementations: A Qualitative and Quantitative Analysis. In Proceedings of the 2023 IEEE International Black Sea Conference on Communications and Networking (BlackSeaCom), Istanbul, Türkiye, 4–7 July 2023; IEEE: Piscataway, NJ, USA, 2023.
24. iPerf—The Ultimate Speed Test Tool for TCP, UDP and SCTP. Available online: https://iperf.fr/ (accessed on 16 October 2023).